  \documentclass[prl,aps,twocolumn,showpacs]{revtex4}
\usepackage[dvips]{graphicx}
\usepackage{dcolumn}%
\usepackage{amsmath}%
\usepackage{amsfonts}%
\usepackage{amssymb}
 \usepackage[usenames]{color}
\providecommand{\U}[1]{\protect\rule{.1in}{.1in}}
\newcommand{\be}{\begin{equation}}
\newcommand{\ee}{\end{equation}}
\newcommand{\bea}{\begin{eqnarray}}
\newcommand{\eea}{\end{eqnarray}}
\newcommand{\bt} {\begin{tabular}}
\newcommand{\et} {\end{tabular}}
\newcommand{\nn}{ \nonumber}

\newcommand{\ba} {\begin{array}}
\newcommand{\ea} {\end{array}}
\begin{document}

\title{Length-dependent thermopower of single-molecule junctions}

\author{  Natalya A. Zimbovskaya}

\affiliation
{Department of Physics and Electronics, University of Puerto 
Rico,  Humacao, Puerto Rico 00791, USA}

\begin{abstract}
In the present work we theoretically study the length dependence of thermopower of a single-molecule junction with a chain-like molecular bridge of an arbitrary length using a tight-binding model. We analyze conditions bringing a nonlinear growth of the thermopower accompanying the  extension of the bridge length. Also, we show that the thermopower may decrease with increasing molecular length provided that the molecular bridge is sufficiently long. 
		\end{abstract}


\date{\today}
\maketitle

{\it I. Introduction:}
 As known, tailored nanostructures hold promise for enhanced efficiency of heat-to-electric energy conversion. Therefore,  thermoelectric properties of tailored nanoscale systems including carbon-based nanostructures, quantum dots and single-molecule junctions have been explored both theoretically and experimentally \cite{1,2,3,4}. Numerous works were focused on Seebeck effect which is directly responsible for conversion of heat to electric energy. This effect occurs when a thermal gradient is applied across a system inducing a current of charge carriers. Seebeck effect in nanoscale systems is measured by recording the voltage $ \Delta V $ which cancels the thermally induced current provided that the temperature difference $ \Delta T $ between two ends of the system is kept constant \cite{5,6}. When $ \Delta T \ll T\ (T $ being an average temperature characterizing the system) the system operates within the linear response regime, so $ \Delta V = - S\Delta T. $ The coefficient of proportionality which appears in this expression is commonly called thermopower. 
 
Various properties of thermopower of single-molecule systems were intensively studied including inelastic effects\cite{7,8,9}, effects of molecular bridge geometry \cite{10,11}, of Coulomb interactions between electrons on the bridge \cite{12,13,14,15}, of molecular vibrations \cite{16,17,18} and of quantum interference \cite{19}. In particular, it was demonstrated that both thermopower and electron conductance of single-molecule junctions may depend on the molecular linker length. Length-dependent conductance and thermopower are usually observed in junctions where the molecular bridge is a chain-like structure consisting of several identical units (e.g. benzene rings) \cite{5,6,11,20,21,22,23,24}.  These molecular linkers provide a better opportunity to observe the relationship between the thermopower and the length of the linker. For other kinds of linkers this relationship is less distinct due to the diversity of specific properties associated with  different parts of the linker.  
 In the most of experiments concerning the issue, the thermopower  appeared to be proportional to the molecular bridge length. However, this linear relationship between the thermopower and the bridge length is not a universal one. Recent experiments carried on single-molecule junctions with gold electrodes and oligophenyl and alcane chain-like linkers showed a distinctly  nonlinear length dependence of the thermopower \cite{23}. The purpose of the present work is to theoretically analyze the possible origin of nonlinearities in the length dependence of thermopower. Also, we discuss specific features of the thermopower versus length profiles.

In the following analysis we assume coherent electron tunnelling to be a  predominant transport mechanism. At  low temperatures the thermopower and conductance through the junction may be computed from the electron transmission $ \tau(E) $ provided that the latter smoothly varies for $ |E - E_F| < kT\ (E_F $ being the chemical potential of electrodes in the unbiased junction) and that the temperature difference between the electrodes $ \Delta T $ is much smaller than $ T $. The corresponding approximations have the form \cite{25}:
\begin{align}
G & = \frac{2e^2}{h}\tau(E_F) \equiv G_0\tau (E_F), \label{1}
\\
S & = - \frac{\pi^2k^2 T}{3|e| \tau(E_F)} \frac{\partial\tau (E)}{\partial E} \Big|_{E = E_F}
\equiv
- S_0 \frac{\partial\ln \tau(E)}{\partial E} \Big|_{E = E_F}.  \label{2}
\end{align}
     
These approximations are commonly used to describe conductance and thermopower dependences on the molecular length and geometry observed in experiments. To employ them  one needs to compute the electron transmission function through the considered molecular junction. Often, $ \tau(E) $ is obtained basing on electronic structure calculations carried out within the density functional theory (DFT) (see e.g. Refs. \cite{10,20,21,22,23,26}). Nevertheless, length-dependent electron transmission may be qualitatively analyzed using simplified Lorentzian and tight-binding models which were invented to explain specific experimental results \cite{20,21,22,23}. 

{\it II. Model and results:} In this work, we simulate a chain-like linker in a single-molecule junction by a periodical chain including $ N $ identical sites. Each site is assigned a single on site energy  $ E_i $ and coupled to its nearest neighbors. For all sites except terminal ones the energies $ E_i $ are supposed to be equal $(E_i = E_0,\ 2 \leq i \leq N - 1) $ whereas $ E_1 = E_N = \epsilon. $ The terminal states are coupled to electrodes through imaginary self-energy terms $ - i\Gamma/2 $ which are supposed to be energy-independent. Also, we assume that the terminal states are coupled to their neighbors in the chain through the coupling parameter $ \delta $ which may differ from the parameter $ \beta$ characterizing the coupling between the remaining sites. The terminal sites are separated out because they may play a significant part in the origin of nonlinearity in the length dependence of thermopower in certain single-molecule junctions. This model is physically relevant for molecular bridges where $\pi-\pi$ dominates electron transport. Then, the parameter $ \beta $ characterizes the coupling between $ \pi $ orbitals  \cite{27}.

Within the accepted model, $\tau(E) = \frac{\Gamma^2}{4} |G_{1N}(E)|^2 $ where $G_{1N}$ is the corresponding matrix element of the retarded Green's function for the chain:
\be
G = (E - H - i\Gamma)^{-1}   \label{3}
\ee
and the Hamiltonian $ H $ is represented by $ N \times N $ matrix:
\be 
H =  \left (\ba{cccccc}
\epsilon - \frac{i\Gamma}{2} & \delta & 0 & 0 & \dots & 0
 \\
\delta & E_0 & \beta & 0 & \dots & 0
\\
0  & \beta & E_0 & \beta & \dots & 0
\\
\dots & \dots & \dots & \dots & \dots & 0
\\
0 & 0 & \dots & \beta & E_0 & \delta
\\
0 & 0 & \dots & \dots & \delta & \epsilon - \frac{i\Gamma}{2}
\\ 
\ea \right ) .  \label{4}
\ee
Using Eqs. (\ref{3}),(\ref{4}) a close expression for $ G_{1N} (E) \  ( N \geq 3) $ may be derived in the form consistent with previously reported results \cite{28,29}: 
\be
G_{1N} (E) = \frac{\delta^2 \beta^{N-3}}{\tilde \Delta_N (E,\Gamma)}. \label{5}
\ee
Here
\begin{align}
\tilde \Delta(E,\Gamma) = & \Delta_N (E,\Gamma) + (\alpha - \lambda)(\alpha + \lambda + i\Gamma)\Delta_{N-2}(E,0) 
\nn\\ & +
\big [(\beta^2 - \delta^2)(\alpha + \lambda + i\Gamma) - (\alpha - \lambda)(\beta^2 + \delta^2) \big]
\nn\\ & \times
\Delta_{N-3}(E,0) - (\beta^4 - \delta^4) \Delta_{N-4}(E,0)  \label{6}
\end{align}
and the determinant $ \Delta_N(E,\Gamma)$ equals \cite{30}:
\begin{align}
\Delta_N(E,\Gamma) = & \frac{1}{2^{N+1} \zeta}\big[(\lambda  + \zeta)^{N-1} (\lambda + \zeta + i\Gamma)^2
\nn\\ & -
(\lambda - \zeta)^{N-1} (\lambda - \zeta + i\Gamma)^2 \big]  . \label{7}
\end{align}
Other determinants included into Eq. (\ref{6}) are given by expressions similar to Eq. (\ref{7}) where $ \Gamma = 0.$ As follows from Eq. (\ref{7}), $ \Delta_0(E,0) = 1 $ and $ \Delta_{-1}(E,0) = 0.$ Also, in these expressions $ \alpha = E - \epsilon,\ \lambda = E - E_0 $ and $ \zeta = \sqrt{\lambda^2 - 4\beta^2}, $ respectively.

Thermally induced charge carriers travel between electrodes using the highest occupied molecular bridge orbital (HOMO) or the lowest unoccupied orbital (LUMO) as transport channels. 
We remark that considering electron transport via HOMO, one should take into account that electron tunneling into a molecule causes an expansion in the eigenstates. As a result, other eigenstates besides HOMO may contribute to the transport even when it occurs at the HOMO resonant energy. However, it does not change the present results (\ref{6}), (\ref{7}) which remain valid.
When the energy values associated with HOMO/LUMO $ E_{H,L} $ noticeably differ from $ E_F\ (|E_{H,L} - E_F| \gg \beta), $ one may put $ \beta = 0 $ in the expression for $ \tilde \Delta_N(E_F).$ This results in the exponential decrease in electron conductance as a function of molecule length typical for off-resonant tunnelling. Such electron conductance behaviour was observed in several experiments \cite{9,20,22,24}. Presenting the conductance in the form $ G(E) = A(E) \exp[-\eta(E)N] $ and substituting this expression into Eq. (\ref{2}), one obtains \cite{21}:
\be
S = - S_0 \left \{\frac{\partial \ln A(E)}{\partial E} \Big|_{E=E_F} - \frac{\partial \eta(E)}{\partial E} \Big|_{E=E_F} N \right\}.  \label{8}
\ee
So, the exponential decrease in the electron conductance  accompanying the increase in the molecular bridge length leads to a linear relationship between the thermopower and the bridge length often reported for single-molecule junctions  with chain-like bridges \cite{9,20,21,22}. Within the accepted model, this conclusion is illustrated in Fig. 1. The electron transmission curves plotted here are computed using Eqs. (\ref{5})-(\ref{7}).Values of  relevant parameters  are chosen to describe electron transport via HOMO. All peaks in the electron transmission function (including that one corresponding to HOMO) are located rather far away from $ E = E_F $ indicating an off-resonant tunneling. Accordingly, both $ S $ and $ \log G $ are linear functions of number of sites $ N, $ as shown in the Fig. 1 (right panel).

\begin{figure}[t] 
\begin{center}
\includegraphics[width=8.8cm,height=4.4cm]{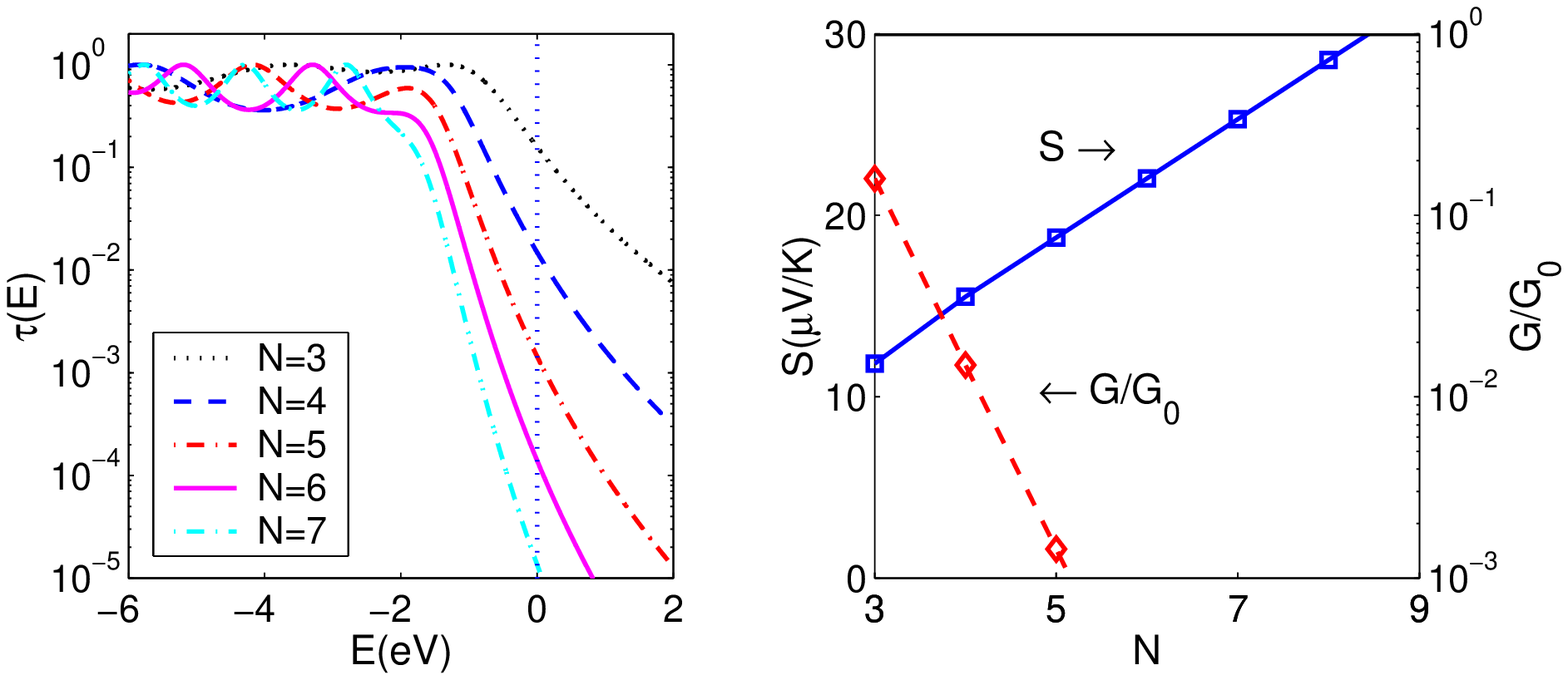}
\caption{ Transmission curves (left panel) and length-dependent thermopower and electron conductance (right panel) computed using Eqs. (\ref{5})-(\ref{7}) assuming that $ E_F = 0,\ E_0 = -4.47 eV,\ \epsilon = -3.82eV,\ \Gamma = 2.86eV,\ \delta = 2.28eV,\ \beta = 1.27 eV,\ kT = 0.026eV.$ 
}
 \label{rateI}
\end{center}\end{figure}

Nonlinear length dependences of the thermopower may appear when HOMO/LUMO is located near $ E_F\ (|E_F - E_{H,L}|< \beta)>$ It was suggested \cite{23} that this may happen
 due to the effect of gateaway states representing bonds between terminal carbons in the bridge and electrodes. Within the accepted tight-binding model, these states are associated with the terminal sites at the ends of the chain. Due to the presence of these states the electron transmission profiles may be significantly affected. At certain values of energies $\epsilon,E_0 $ and coupling parameters $ \delta $ and $ \beta ,$ peaks associated with HOMO in the plots of $ \tau (E) $ versus $ E $ become broader than other resonance features, as shown in the left panels of Figs. 1,2. When these distorted HOMOs are located near $ E = E_F, $ the thermopower  displays a nonlinear dependence on $ N $  presented in Fig. 2. We remark, that the values of all relevant energies used in computations of the electron conductance and thermopower displayed in this figure are the same as those used in Ref. \cite{23}, so the results for $ G(N) $ and $ S(N) $ at  $ 3 \leq N \leq 6 $ agree with the corresponding results presented in that work. One observes that nonlinear length dependence of the thermopower  (and, to a smaller extent, of $\log G )$ is noticeable while the bridge is rather short. For $ N \geq 6 ,$ both $ S $ and $ \log G $  become linear functions of the bridge length. This happens because the effect of gateaway states on the electron transmission is fading away as the bridge   lengthens  and HOMO moves away from $ E = E_F. $

\begin{figure}[t] 
\begin{center}
\includegraphics[width=8.8cm,height=4.4cm]{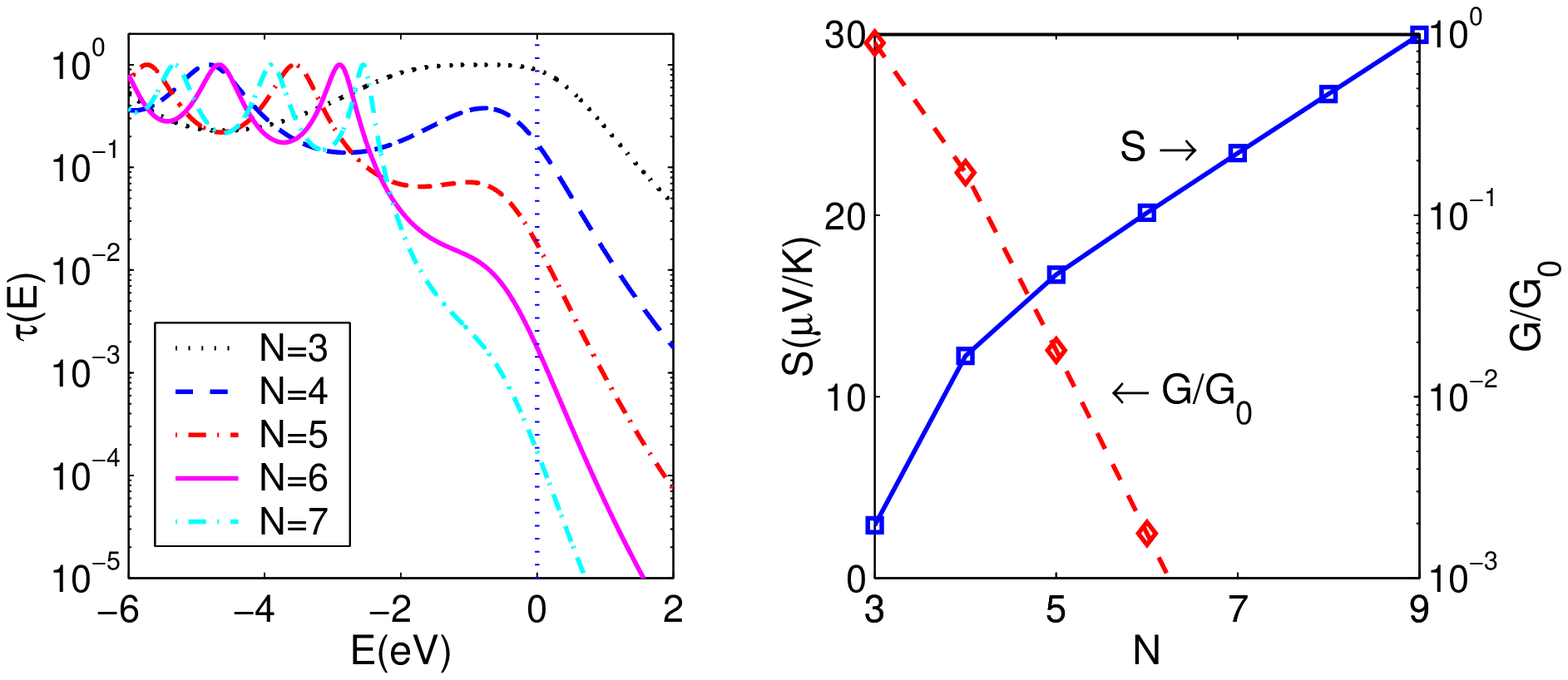}
\caption{Transmission curves (left panel) and length-dependent thermopower and electron conductance (right panel) computed using Eqs. (\ref{5})-(\ref{7}) assuming that $ E_F = 0,\ E_0 = -4.47 eV,\ \epsilon = -1.85eV,\ \Gamma = 2.86eV,\ \delta = 2.28eV,\ \beta = 1.27 eV,\ kT = 0.026eV.$ 
}
 \label{rateI}
\end{center}\end{figure}  

As known, only HOMO/LUMO participate in thermally induced charge transport through a single-molecule junction provided that a temperature gradient applied across the system is sufficiently small. This gives grounds to disregard all molecular resonances except those corresponding to HOMO/LUMO thus reducing the electron transmission to a single peak nearest to $ E = E_F. $ Lorentzian models were used to represent this peak in the  computations of the junction thermopower  in several works\cite{6,21,31}. To describe the length-dependent thermopower, the coupling parameter between the bridge represented by a single orbital and the leads should be simulated by a length-dependent function. This function should be chosen in such a way that the peak associated with HOMO/LUMO would become  sharper and narrower as the bridge length increases. Such models may predict nonlinear dependences of the junction thermopower on the bridge length. For example, the following model for the electron transmission \cite{31}:

\be
\tau(E) = \frac{\Gamma^2}{4} \frac{\delta_N^2}{\big|(E - E_{H,L} - i\Gamma/2)^2 - \delta_N^2\big|^2}  \label{9}
\ee
where $ \delta_N = \delta_0 \exp[-\gamma N]\ (\delta_0 $ and $ \gamma $ being independent on energy and length) brings the expression for the thermopower which saturates for long molecular bridges approaching  the limit:
\be
S_\infty = \frac{4S_0(E_F - E_{H,L})}{(E_F - E_{H,L})^2 + \Gamma^2/4}.  \label{10}
\ee

\begin{figure}[t] 
\begin{center}
\includegraphics[width=8.8cm,height=4.4cm]{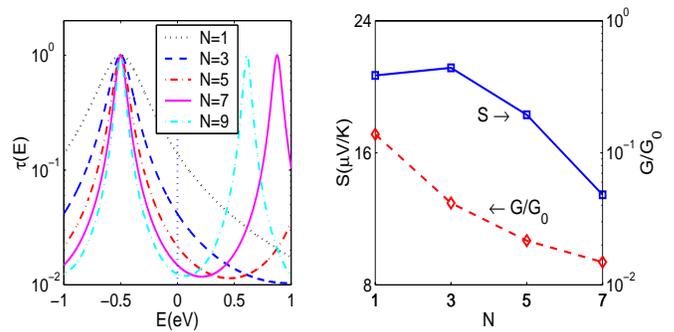}
\caption{Transmission curves (left panel) and length-dependent thermopower and electron conductance (right panel). The curves are plotted assuming  $ E_F = 0,\ E_0 = \epsilon = -0.5 eV,\  \Gamma = 0.2 eV,\ \delta= \beta = 1.28eV,\  kT = 0.026eV.$ 
}
 \label{rateI}
\end{center}\end{figure}  

Within the tight-binding model adopted in the present work, one observes that the electron transmission function for the chain including an odd number of sites always shows a Lorentzian-like peak at $ E = E_0 ,$ (see Fig. 3). As the number of sites increases, the width of this peak shrinks although its height remains the same. Assuming that this peak is located slightly to the left from $ E = E_F $ and may  serve as HOMO, we compute the molecular conductance and thermopower of the corresponding system using Eqs. (\ref{1}),(\ref{2}). The results are presented in the right panel of Fig. 3. As shown in this figure, the thermopower may reach a maximum at a certain bridge length, and then decrease as the bridge further extends. This is accompanied by a pronounced nonlinearity in the length dependence of $ \log G. $ This behavior of the thermopower originates from alterations in the transmission peak shape occurring as the number of sites in the bridge increases. At sufficiently small values of $ N ,$ the slope of the curve $ \tau(E) $ at $ E = E_F $ grows as $ N $ enhances. However, at greater values of $ N $ the slope significantly decreases causing the thermopower to decrease. We remark that the decrease in thermopower accompanying the extension of the molecular linker length was not reported so far, although some data experimentally obtained for alcane chains linking gold electrodes \cite{23} may indicate that the thermopower of these molecular junctions could either saturate or decrease for longer chains.

{\it III. Conclusion:} 
In the present work we have employed a tight-binding based model to describe a single-molecule junction with a chain-like bridge consisting of an arbitrary number of identical units. We used this model to qualitatively analyze the length-dependent thermopower of the considered system. The obtained results confirm that the character of thermopower dependences on the molecular length is determined by the location of HOMO/LUMO with respect to the chemical potential of electrodes and by the profile of the corresponding peak in the electron transmission function. It is shown that characteristics of terminal sites coupled to the electrodes may bring a nonlinear length dependence of the thermopower when the bridge is sufficiently short. This conclusion agrees with the results reported in Ref. \cite{23}. Also, it is predicted that under certain conditions the thermopower of a single-molecule junction may decrease as the molecule length increases. Since experiments on the thermopower of molecular junctions are often carried out at room temperature, finite-temperature effects may play a part in controlling the length dependences of the thermopower. They can affect it by giving rise to fluctuations in the molecular bridge geometry and to molecular vibrations \cite{6,7,32,33}. The model employed in the present work may be generalized to include electron-vibron interactions following the way suggested in Ref. \cite{34}.  
It may be further improved by including into consideration the energy shift which appears due to the real parts of self-energy terms describing the coupling of the molecular bridge to the leads. There are some grounds  to conjecture that this energy shift may influence the thermopower behavior.
Finally, the analysis could be extended to study length-dependent thermoelectric properties of single-molecule junctions beyond linear in $ \Delta T $ regime. However, these studies are beyond the scope of the present work. Nevertheless, despite the simplicity of the adopted model  
we believe that presented results may help in further understanding of thermoelectric properties of single-molecule junctions.

 {\it Acknowledgments:}
The author  thank  G. M. Zimbovsky for help with the manuscript. This work was supported  by  NSF-DMR-PREM 1523463.

\end{document}